# Developing a Methodology for Online Service Failure Prevention: Reporting on an Action Design Research Project-in-Progress

**Jacques Louis Du Preez**
School of Information Management
Victoria University of Wellington
Wellington, New Zealand
Email: duprejacq@myvuw.ac.nz

**Mary Tate**
School of Information Management
Victoria University of Wellington
Wellington, New Zealand
Email: mary.tate@vuw.ac.nz

**Alireza Nili**
School of Information Management
Victoria University of Wellington
Wellington, New Zealand
Email: alireza.nili@vuw.ac.nz

## Abstract

The increasing use of online channels for service delivery raises new challenges in service failure prevention. This work-in-progress paper reports on the first phase of an action-design research project to develop a service failure prevention methodology. In this paper we review the literature on online services, failure prevention and failure recovery and develop a theoretical framework for online service failure prevention. This provides the theoretical grounding for the artefact (the methodology) to be developed. We use this framework to develop an initial draft of our methodology. We then outline the remaining phases of the research, and offer some initial conclusions gained from the project to date.

## 1. Introduction

Services are becoming an increasingly large and important sector of the world economy, and are estimated to make up approximately two thirds of the total world Gross Domestic Product (GDP) according to the World Bank (World Bank 2012). Over the years the nature of services and service delivery has changed substantially, with services becoming increasingly digitalised. Digital services enable customers to interact with a user interface (e.g. a mobile phone) in order to achieve a desired benefit (Fassnacht & Koese 2006). Digital services are increasing in both the extent of their use, and the importance and complexity of the transactions conducted. For example, it has been estimated that 46% of service transactions conducted with the New Zealand government are carried out online (up from 30% in 2012)[1]. This now includes essential services such as registering births and applying for passports[2]. It is not only stressful and inconvenient if these services are not carried out correctly, but in some cases satisfactory completion of the transaction is a legal requirement (e.g. registration of birth). Therefore, preventing digital service breakdowns is becoming increasingly important for service providers and users (Tate et al. 2014; Tate & Evermann 2009; Tate & Evermann 2010).

---

[1] https://www.realme.govt.nz/news/kiwis-big-users-online-services/
[2] https://www.realme.govt.nz/news/kiwis-big-users-online-services/

Services have a number of different stakeholders, with varying goals and objectives. According to White and Leifer (1986, p. 215) "perceptions of a systems' success or failure may vary depending upon an individual's perspective of the system". Customer's perceptions of online service failure might not always align with the online service provider's perception of online service failure. Furthermore, in an online environment, organizations may not receive direct feedback from customers. As a result service providers can be blissfully unaware that online service failure has occurred. This is dangerous, since service failures can result in significant losses to firms, through negative word of mouth and lost customers (Bitner et al. 2000). Service failure is also a key driving factor behind customer switching behaviour (McCollough et al. 2000). This is especially relevant to online services, due to the 'relative ease of switching over to another service provider. Therefore online services and self-services should concentrate on preventing failures from occurring in the first place, as well as identifying and recovering from failures that do occur.

This work-in-progress action design research aims to develop a methodology to assist organisations and their customers to prevent online service failures from occurring. The objective of this methodology is for it to be usable by organisations to reduce failure as well as improve their client's customer experience.

We first review the theoretical foundations of service failure prevention and service management methodologies. Next we introduce the action design research (ADR) methodology (Sein et al. 2011). We report the outcomes of the first stage of our ADR, which is the development of an alpha version of our methodology, based on theory. Finally we describe the plan for the remainder of the research, and offer a brief conclusion.

## 2. Literature Review

We first explain the concept and causes of online service failure. We then consider strategies for service failure prevention. Assuming that failures will sometimes still occur, we also look at service recovery, and how insights from service recovery can be captured to prevent future failures. These streams of literature are integrated into a service failure prevention framework. Then we briefly investigate existing service management methodologies and show where our contribution is positioned.

### 2.1 Online services, service failure and service quality

Traditionally service definitions concentrated on the differences between products and services. The "IHIP" conceptualization of services defines them as Intangible, Heterogeneous (each instance of delivery is different), Inseparable (production and delivery are inseparable) and Perishable (Zeithaml et al. 1985). However online services are normally standardised, not heterogeneous, tangible, non-perishable, and developed independent of their consumption (Tate & Evermann 2010), so a new definition is required. We use Fassnacht and Koese's (2006) definition of an online service: "Services delivered via information and communication technologies where the customer interacts solely with an appropriate user interface (e.g., website or online application) in order to retrieve desired benefit" (Fassnacht & Koese 2006, p. 23). As a result online services also include E-commerce activities, which is defined as "the use of the internet to facilitate, execute, and process business transactions" (Delone & Mclean 2004).

The concept of online service failure is based on Expectation-Confirmation Theory (ECT) (Oliver 1980) and is defined based on the traditional "gap" model of service quality (Parusaraman, Zeithaml & Berry 1985). Online service failure is "the gap that occurs when customers perceived quality of service delivery does not match their service expectations" (Nili et al. 2014, p. 2). Forbes et al. (2005) also highlight that e-tail customers experience different failures in relation to their traditional retail counterparts. As a result new failure-recovery and -prevention strategies are required to address online service failures.

Understanding service quality is another key concept that needs to be kept in mind when thinking about service failure prevention, since a customer perception of poor quality represents a degree of failure, and areas that need to be improved. One of the influential studies of online service quality, by Zeithaml et al. (2000) identified 11 e-service quality dimensions relating to websites, including: reliability, responsiveness, access, flexibility, ease of navigation, efficiency, assurance/trust, security/privacy, price knowledge, site aesthetics and customization/personalisation (Zeithaml et al. 2000). These dimensions can be used to assess the service quality of an organisations online services. These dimensions can also point out service failures that need to be addressed to improve the organisation's online service quality. Other sets of quality dimensions that could be referred to for insights include those developed by Fassnacht and Koese (2006); Baierova Tate & Hope (2003); Barnes and Vidgen (2002).

## 2.2. Causes of service failure

We use the detailed set of online service failure categorisations and definitions developed by Tan et al. (2011) to guide the research. According to this categorisation there are three types of service failure associated with online services, namely: Informational failures, Functional failures, and System failures. Functional failure refers to online service failures where the provided functionalities are insufficient or unable to support customers in the accomplishment of a transaction, resulting in failure in terms of meeting the user's functional requirements/expectation. For example, a banking system that does not provide notification when a customer is going into overdraft and accumulating penalty interest might be working "as specified" but not meet the customer's service expectations. Informational failure refers to online service failure that occurs as a result of customers receiving irrelevant, inconsistent or incomplete information that negatively impacts their service experience. An example would be advising online that a product is in stock, only to find when trying to order one, or when going to a store to buy one that it is not available. System failure refers to online service failure that is caused by functionalities of the website not being delivered properly, resulting in a negative experience for the customer. This includes a wide range of technology failures, including hardware, software, and network issues. These failure types are not mutually exclusive. Nili et al. (2014) explain that informational and system failures can lead to functional failures as well.

## 2.3. Online service failure prevention

Effective failure prevention requires an in-depth understanding of the stages of service delivery, the risks and vulnerabilities at each stages, and the techniques available to prevent them.

Sauer (1993) suggests organisations should conceive failure as a process, rather than a single discrete event. The service delivery process is also composed of a number of different stages, relying on each other directly or indirectly. Therefore the service value-chain framework developed by Nili et al. (2014) provides a useful basis for understanding and analysing service failure prevention, as different types of failure may be identified at different stages of the value chain (Figure 1).

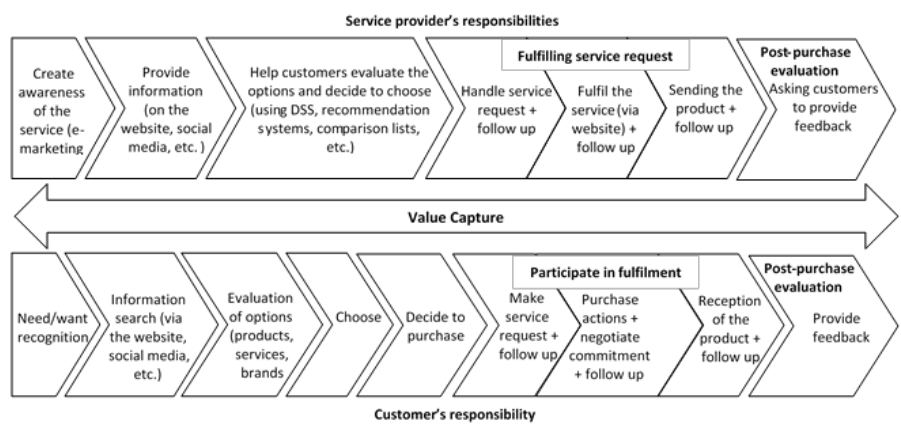

*Figure 1: Digital service value chain framework (adapted from Nili et al., 2014)*

Risks and vulnerabilities in online service delivery can arise from multiple points in the value chain, and if a service failure occurs at one stage in the process it can have flow on effects, which in turn can cause failures at other points (Halstead et al 1996; Chuang 2007). Song et al. (2013) suggest organisations conduct a risk assessment of different types of failures, considering the *probability* of service failures (PSFs) as well as assessing the *impact* of service failures (ISFs). This information should then be used to develop a service failure assessment map. This is a useful approach to prioritise which aspects of the online service is most vulnerable to failure, highlighting areas that will benefit the most from the development of failure prevention strategies. Some of the identified vulnerabilities might be easily addressed by implementing technology enablers.

A risk analysis of service failure needs to encompass both the organizational and the customer perspective (Nili et al. 2014). When customers encounter a problem or annoyance it can lead to adverse effects such as motivating them to switch to competitors or engage in negative word of mouth, which can have a serious impact on the organisation (Stauss 1993). Chandani and Gupta (2014) highlight the importance of addressing failures, and stress the importance of identifying the root cause of the failure, so that it will not reoccur in the future. They provide a logical defect prevention flow model outlining that failures need to be identified, logged and classified. After this a root cause analysis needs to be performed to identify the cause of the failure. Once the cause has been identified corrective action can be applied to resolve the issue. Understanding how to resolve the issue can inform the development of failure prevention strategies to reduce the likelihood of the failure reoccurring in the future.

### 2.3.1. Technologies for service failure prevention

Improving processes to prevent online service failure is not the only way to address online service failure concerns. Effective use of preventative technologies at appropriate points in the service delivery process can prevent online service failures. Technology enablers, in this context, refer to "specific technologies and technological approaches that can be used to prevent website service failure" (Nili et al. 2014, p. 6). For example, technology enablers such as online chat capabilities, social media, online tutorials, and automated messages can be utilised to interact and communicate with customers to prevent, or at least reduce the likelihood of service failures occurring (Nili et al 2014). Other examples include network intrusion detection software, firewalls, VPNs, proxy servers and network address translation to prevent system failures. Nili et al. (2014) classified service failure prevention technologies according to the type of failure that the technology addresses (using the system failure, information failure and functional failure classifications from Tan et al. (2011)). They also classified prevention technologies based on the points in the digital service delivery value chain at which they can be applied.

The way an organisation designs the interface of online services can also have an impact on preventing certain kinds of failure. Kirakowski et al. (1998) identified four factors that influences a customer's perception of a website, including: attractiveness, control, efficiency and helpfulness. Usability is another key aspect of online services and can be analysed through heuristic evaluations, cognitive walkthroughs and user testing to identify usability problems (Chen & Macredie 2005). Among these techniques, heuristic evaluation is the quickest, cheapest and most effective way to identify usability problems (Greenberg et al. 2000). Chen and Macredie (2005) developed usability guidelines based on Nielsen's heuristic evaluation (Nielsen 1994). Their usability guidelines outlines a range of interface considerations that need to be taken into account when developing a website which will enable a customer to use it effectively and minimise perceived online service failures.

## 2.4. Recovery strategies associated with online services

Even with the best prevention, online service failures cannot always be anticipated. This means that learning from past service recovery events is also an important component of prevention. Effective service recovery offers opportunities for improving a service in the future (Spreng et al. 1995). Improvement to the online service can lead to both cost reductions, through changing inefficient and ineffective processes, and by reducing the likelihood of future failures from occurring, and thus leading to fewer dissatisfied customers (Johnston & Michel 2008).

Michel et al. (2009) argue that service recovery often fails as a result of unresolved tensions between customer recovery, process recovery and employee recovery. Johnson and Michel (2008) highlight the intended outcomes of each of these recovery types. The purpose of customer recovery is to return the customer to a satisfied state, while the purpose of process recovery is to learn from the failure and to make the necessary improvements to the process to reduce the likelihood of the failure reoccurring. The role of employees in digital service recovery is often overlooked. Johnson and Michel (2008) came up with seven key activities involved in customer recovery after a service failure has occurred. These activities include: acknowledgement, empathy, apology, own the problem, fix the problem, provide assurance and provide compensation (Johnston & Michel 2008). Process recovery, on the other hand, moves the focus away from individual customers, towards management activities that improve the processes associated with the service delivery to ensure future customers are satisfied (Johnston & Michel 2008). According to Johnston and Clark (2005) process improvement involves four key stages, including: data collection, data analysis, costing and improvement. As a result organisations will have to go through these stages to recover their processes after an online service failure has occurred to prevent it from happening again. The purpose of employee recovery is to support staff through failures, by providing them with training and rewards for dealing with failures, since they are often the ones that have to deal with the fallout of the failure, negatively impacting their motivation. Michel et al. (2009) state that the integration of these different perspectives of service recovery is required to successfully recover from service failure.

### 2.4.1. Culture of learning

Learning from service recovery and implementing changes is essential to prevent future online service failures from occurring. Edmondson (2011) suggests that organisations should develop a culture of learning, where organisations strive to understand why failure occurred and not who is responsible. For this approach to work, consistent reporting of failures, large and small, combined with systematic analysis is required. Through the development of a culture of learning organisations will be more accepting of hearing about failure, since it will enable the organisation to improve its service and prevent the same failures from reoccurring.

### 2.4.2. Capturing the customer perspective on online service failure

Wilson and Howcroft (2002) stressed the importance of understanding others interests and aligning it with your own. Aligning the various stakeholder's interests and objectives as much as possible will enable a move towards a shared understanding of success and failure (Wilson & Howcroft 2002). One way in which this can be achieved is through providing customers with adequate channels through which they can make complaints about the service quality or where they can provide suggestions as to how the online service can be improved. Research has shown that harnessing customer's feedback given by willing consumers regarding their negative experiences provides firms with an invaluable source of information (Blodgett & Anderson 2000).

In conclusion, an integrated conceptual framework for service failure prevention, showing the relationships between the theoretical components we have discussed, is shown as Figure 2. We show that capturing the perspectives and expectations of stakeholders at each stage in the value chain is essential for understanding online service failure. Once expectations are understood, a risk analysis should be carried out to identify high risk and high impact failures and their root causes. Strategies can then be developed to prevent these. These may include the deployment of appropriate technologies, as well as aspects of the interface and information design. This needs to be accompanies by a culture of organizational learning which captures insights from those failures that still occur.

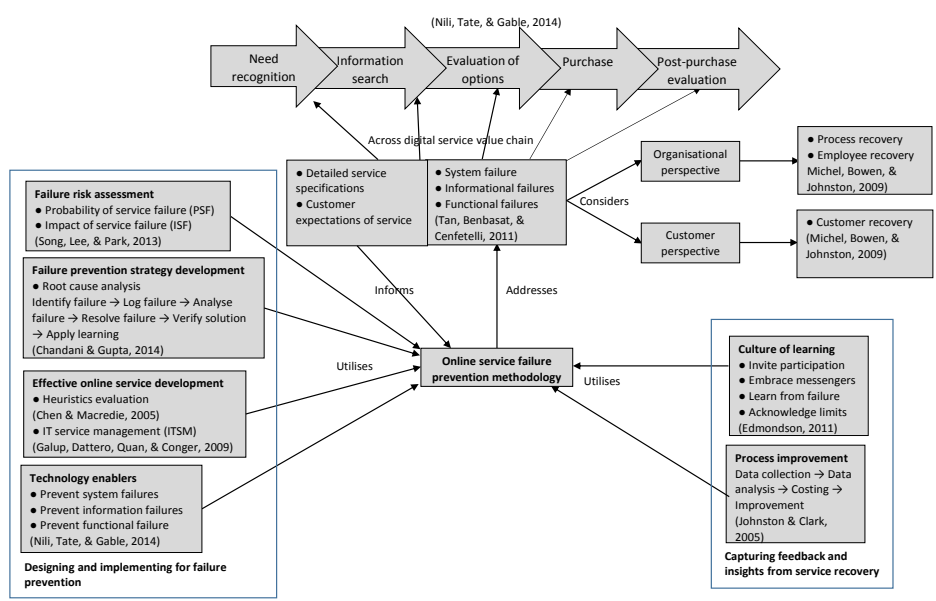

*Figure 2: Integrated Framework for Service Failure Prevention*

## 2.5. Methodologies for service management

We note that our method assumes that an initial, detailed specification of the service requirements and service levels at each phase of the value chain will be an *input* to the failure prevention method. It is necessary to have a clear specification of what constitutes acceptable service (informational, functional and technical) before failure prevention can be considered.

A further important consideration is capturing the input and expectations of the customer during the service specification and design process. We have suggested that situations the customer perceives as service failures may go unobserved by the organization. Particularly in self-service contexts, the immediate feedback obtained from direct interactions with customers is not provided by digital service delivery. We have emphasized the importance of "embracing messages" once a service has been delivered, but it is equally important to understand customer perceptions of success and failure as deeply as possible in advance of initial service implementation.

We also need to position our methodology development within the context of existing best practice frameworks. Best practices associated with online service management are articulated in IT service management (ITSM) frameworks, which draw, in turn, on other quality management frameworks such as TQM, business process management, Six Sigma and CMMI (Galup et al. 2009). ITSM is often associated with the British Government's Information Technology Infrastructure Library (ITIL). ITIL is a framework of best practises intended to facilitate the development and delivery of quality IT services, built around a process-based systems perspective of managing IT operations, including continual improvement and metrics (Galup et al. 2009). The 3rd version of ITIL consists of five publications addressing: service strategy, service design, service transition, service operations, and continual service improvement (Taylor 2007). The ITIL framework, however, does not specifically address failure prevention.

We note that some skills-sets, such as risk management, or user experience design have their own knowledge areas and methodologies. These are often used in conjunction with broader project management or system development methodologies. We therefore intend, in this research, to develop a specialized approach for failure prevention that will complement other methodologies that organizations may be using, and that can be customised and integrated with existing organizational methodologies.

In our approach to methodology development, we draw on leading system development and project management methodologies, modified for our purposes. The classic "waterfall" system development life-cycle includes: analysis of requirements, design, construction, testing and ongoing system support (Schwalbe 2014). From this, the analysis and design phases are most relevant. Failure prevention strategies need to be planned and analysed, drawing on the value chain stages, stakeholder perspectives, risk management approaches, root cause analysis techniques, and technologies for failure prevention that we discussed above. Prevention strategies then need to be designed in detail. Not all prevention strategies will be "constructed", some may be implemented as process improvements. Traditional testing may not be necessary, as some strategies will be evaluated based on the effectiveness in reducing service failure. It is also important to take a "whole of life" view of failure prevention, so rather than "support" as the final life-cycle stage, failure prevention needs to be considered as an ongoing process of learning and improvement which is integrated with failure recovery and organizational learning. The classic project life-cycle phases are considered to be initiating, planning, executing, and closing (Schwalbe 2014) with monitoring and controlling as ongoing activities throughout the project life-cycle. From this, we include monitoring, controlling and ongoing evaluation processes.

In summary, we propose a service failure prevention methodology should include detailed analysis of the risks and root causes of service failure, from the perspective of the organization and the customer, and with consideration of system, functional, and informational failures. Prevention strategies should be designed and developed. Once digital services are implemented, failure prevention needs to be an ongoing process that is integrated with service recovery strategies. Overall, monitoring and controlling digital service failure levels, and implementing strategies to prevent and reduce failures, needs to be integrated into organizational performance and control measures.

## 3. Research Method

In our research we will be following an Action Design Research (ADR) approach (Sein et al. 2011). This combines action research and design research. Action research addresses real world problems in participatory and collaborative ways, with the aim of producing knowledge through a cyclical process (O'Leary 2004). Design research aims to develop prescriptive design knowledge through building, evaluating and improving innovative IT artefacts to solve identified problems (Hevner et al. 2004, March & Smith 1995).

### 3.1. Action design research

ADR reflects the premise that IT artefacts are constructed and shaped by the organisational contexts during its development and use (Sein et al. 2011). The ADR approach allows us to link theory with practice, while still allowing practice to influence the final IT artefact. We will start with a theory-driven artefact, which will be refined in an organizational setting. An overview of the phases of ADR that will be carried out for this project follows.

#### 3.1.1. Problem formulation

Firstly ADR requires an issue or problem to address, which can be triggered by a perceived problem in practise or an anticipated problem by researchers (Sein et al. 2011). In this case we aim to prevent online service failures from occurring.

#### 3.1.2. Building, intervention, and evaluation

In the next stage of the ADR approach we will continue to develop our online failure prevention methodology. This will be accomplished through following an iterative process that includes three phases (Figure 3). The three phases include the Building phase, Intervention phase and the Evaluation phase (BIE), with the outcome of the process being the methodology (Sein et al. 2011).

The building phase involves developing our online service failure prevention methodology. Initially our methodology is based on theoretical understandings of online service failure prevention, turning findings from literature into methods and practices that are useable by organisations. The outcome of this phase is presented here. Once we developed our initial methodology we will move on to the intervention stage, which involves applying the artefact in an organisational contexts (Sein et al. 2011). We will initially workshop and refine our methodology with expert practitioners to develop an alpha version. Following that, the alpha version will be applied to developing and implementing a service prevention strategy in a specific organisational context. In the evaluation phase we will evaluate the quality of our methodology looking at its usefulness, effectiveness, completeness and impact on the organisation (Hevner et al. 2004). This feedback will be gathered qualitatively, through a focus groups which will consist of a number of specialists within the organisation. Feedback provided by the focus group members will enable us to make further improvements and refinements to our methodology. We will also gather quantitative metrics of organizational performance (these will be determined in consultation with the organization, and compare services where the failure prevention method has been applied to those where it has not. After the evaluation stage we will iterate back to complete another BIE cycle.

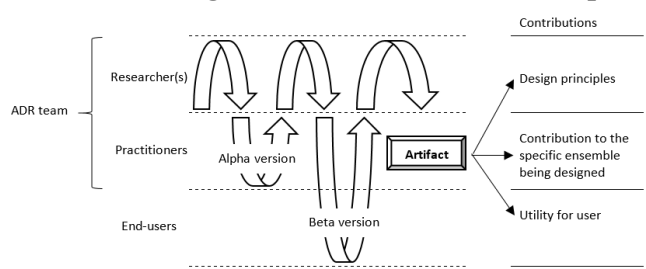

*Figure 3: Generic schema for BIE (adapted from Sein et al. 2011)*

#### 3.1.3. Formalisation of learning

Once we are satisfied with our online service failure methodology we can move on to the formalisation of learning phase. The aim of this phase in the ADR approach is to communicate back the findings of the research to industry, while also formalising the learning and extracting theory. Therefore to formalise the learning of the study we will communicate back our findings regarding the specific-and-unique organisational contexts while also addressing how our methodology can be applied to other organisations that face similar issues (Sein et al. 2011). We will communicate back our research findings through scholarly publications. The first step in socializing the method with the academic community is presented in this work-in-progress paper.

## 4. Results

Based on the framework developed from literature (Figure 2) we developed a first cut version of the methodology. The specification of the service acts as input to the method. We note that this should include understanding of the stages in the value chain, and customer expectations of service levels at each point.

The first stage of the method involves planning and designing to prevent service failures. The inputs to this stage are the service specification, and customer expectations of service levels at each stage in the value chain. These specifications and service levels are subjected to a risk analysis to determine the probability of failure occurring, and the impact of the failure. From this high impact and/or high probability risks can be analysed in more detail using root cause analysis. Based on this analysis, technology and interface design solutions can be developed to prevent high risk failures. The importance of "designing in" failure prevention, rather than relying on failure response and service recovery strategies cannot be over-estimated, and is a key contribution of this study. Once technology architecture and user interface design decisions are implemented they are much more difficult and expensive to change, and failures tend to result in "band-aid" solutions resulting from an unwillingness to engage in large-scale redesign.

The next stage is to implement the failure prevention strategies along with feedback mechanisms. These may include more traditional metrics such as availability, as well as customer feedback, and insights derived from data analytics.

The final stage is ongoing learning to prevent future failures. This stage emphasizes "whole of life" management for the digital service. Recovery strategies, including organizational process improvements, customer actions, and employee actions should be in place. More importantly, the organization needs a culture of learning from failure and capturing and welcoming customer feedback. Without the immediacy of face-to-face customer contact, gradual customer attrition may occur for online services (or the uptake may be lower than expected), which can be difficult for the organization to explain. Capturing customer perceptions of service failure on an ongoing basis is essential for future failure prevention.

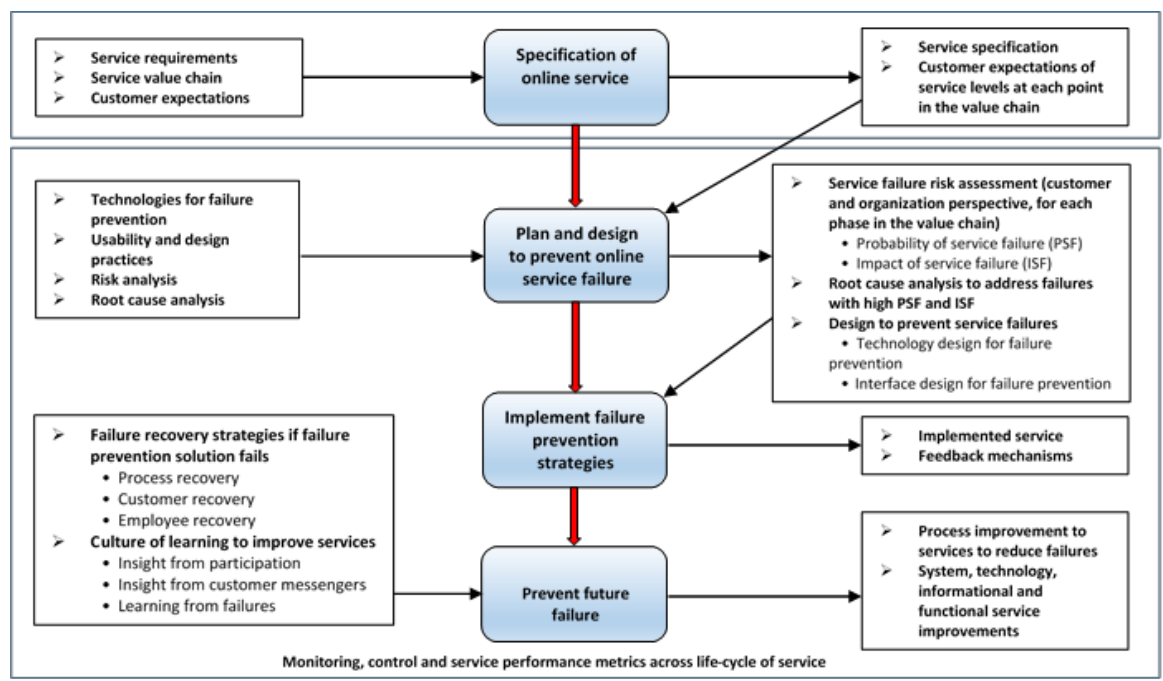

*Figure 4: A first cut methodology for service failure prevention*

Drawing on project management life-cycles, we also suggest that ongoing monitoring, control, and performance management needs to take place throughout the life-cycle of the service.

## 5. Next Steps

As we described in our methodology section, this work-in-progress paper reports the first stage of an action design research project, which is developing a theory-inspired artefact. We conducted a wide-ranging literature review, drawing on theories of services, service quality, service value creation, risk management, interface design, service recovery, and organizational learning to provide an integrated approach to service failure prevention. We then applied these insights, in conjunction with insights about methodologies, to the development of the first draft of our methodology (reported here). In the next steps, we envisage collaborating with a local body organization that is progressively rolling out digital services to rate-payers to carry out the remaining phases of the ADR method.

The next steps of the research will include:

1. Presenting the theoretical foundations and first cut of the methodology in a workshop to service managers, and obtaining detailed feedback. This feedback will inform elaboration and revision of the method (Alpha version).
2. Collaborating with a service manager/designer in an organizational context to design an intervention. The methodology will be applied to a digital service which has been specified, but not yet developed and implemented. Interventions will be planned, and measures of success developed. The methodology will be refined (Beta version).
3. Working with organizations to incorporate and adapt the methodology to their own context and practices so it can be embedded.
4. Ongoing monitoring of performance, in conjunction with the organization, including the effective management of service failure recovery, the culture of organizational learning, and the ability to embrace customer "messages".
5. Reporting the evolving method for practitioner and academic audiences.

## 6. Conclusion

Designing robust online services and preventing expensive failures requires a multi-faceted approach drawing on a range of knowledge areas. Our major initial insights are that 1) prevention needs to be "designed in" from the start, and requires a culture that is prepared to balance the time needed to design for failure prevention with the demands of timely delivery; and 2) a culture of learning and openness, and a willingness to engage with stakeholder feedback is essential if the insights from service recovery are to be captured to prevent further failures.

Achieving these things requires recognition that understanding customer expectations of service levels throughout the phases of the service delivery process is critically important, as not all customers that successfully complete the steps in an online transaction will consider the experience to be a success, while conversely, some that only engage in some early stages on the value chain (for example, searching and evaluation of options) may be very satisfied. We propose that organizations should follow a method specifically aimed at failure prevention, which should be carried out once the initial specification and service levels have been completed. The method should draw on existing knowledge of risk analysis and root cause analysis, complemented by the appropriate use of both technology and interface design solutions to prevent high risk failures. Recognizing that failures will still occur, capturing input from stakeholders, and insights from service recoveries is essential to avoiding recurrence of failures.

While individual organizations have slightly different in-house methodologies for project management, system development, risk management, interface design and other competencies that are related to service development and delivery, these in-house methodologies share common principles from theory. In this paper, we articulate some common principles for online service failure prevention, as the first step in developing a theoretically grounded, usable method for organizations to engage in effective online service failure prevention.